\documentclass[preprint,preprintnumbers,amsmath,amssymb]{revtex4}

\usepackage[dvipdfmx]{graphicx}
\usepackage{amssymb}
\usepackage{color}
\usepackage{ulem}

\newcommand{\nn}{\nonumber}

\begin{document}

\title{Mass generation via nonlinear massive solution in Higgs potential and particle creations}

\author{Yoshio Kitadono}
\email{kitadono@ncut.edu.tw}
\affiliation{Liberal Education Center, National Chin-Yi University of Technology, \\
	No.~57, Sec.~2, Zhongshang Rd., Taiping Dist., Taichung 41170 Taiwan}
	
\author{Tomohiro Inagaki}
\email{inagaki@hiroshima-u.ac.jp}
\affiliation{Information Media Center and Core of Research for the Energetic Universe, Hiroshima University,\\ 
	No.~1-3-2, Kagamiyama, Higashi-Hiroshima, Hiroshima 739-8521 Japan}

\date{\today}

\begin{abstract}
The nonlinear massive plane wave solution of the classical scalar field in the Higgs potential is revisited to study the mass generation and particle creation. In particular, by assuming that the Higgs system is in the slightly excited state in early universe and it is described by the nonlinear solution, we study the mass generation mechanism for massive vector bosons and a heavy fermion in the quantum field theory around the nonlinear massive classical field. The nonlinear massive classical solution gives the transition from the vacuum to a pair of vector bosons and fermions. We present the new formulae of the probability density of the production process for particles in the standard model of elementary particle physics. The probability densities of the particle productions vanish when the nonlinear massive solution reduces to the constant solution (the classical vacuum expectation value); while the probability densities are expressed as the function of the free parameter in the classical solution in general case. We discuss the behavior of the probability densities for the three oscillating modes in the classical solution.
\end{abstract}


\maketitle

\section{Introduction \label{sec1}}
\noindent

\indent
The mass of an elementary particle is one of mystery in physics. In particular, how are the masses of elementary particles generated ? In 1960's, the so-called "Higgs mechanism" \cite{Englert.Brout.1964, Higgs.1964, Guralnik.Hagen.Kibble.1964} was proposed to generate the masses of the gauge bosons by keeping compatibility with gauge theory in the standard model (SM) of physics. As the consequence of the Higgs mechanism, the Higgs particle was predicted and scientists had been investigated to detect this particle for long time. Finally, in 2012, the ATLAS and CMS collaborations announced the discovery of the Higgs boson with $125~\mbox{GeV}$ mass \cite{ATLAS.Higgs.2012, CMS.Higgs.2012}. 

In the Higgs mechanism, the quadratic term of the Higgs field with the negative mass-squared and the quartic self interaction of the Higgs field in the Higgs potential play the crucial role for the mass generation. The quartic interaction in the Higgs potential or Yang-Mills theory gives an anharmonic oscillator as a solution of equation of motion (EOM) and its quantum theory has special importance, for instance, the large order perturbation theory \cite{Bender.1969}, the quantization around nontrivial classical solution and nonpertubative aspects of the theory \cite{Dashen.1974}, and Borel summation \cite{Graffi.1980}. One of the interesting classical solution in Yang-Mills theory which is known as nonlinear plane wave solution described by the Jacobi's elliptic solutions was discussed \cite{Treat.1971}. 

As the existence of the nonlinear plan wave solution was suggested by Coleman \cite{Coleman.1977},  the nonlinear plane wave solutions in $\mbox{SU}(2)$ gauge theory were investigated \cite{Corrigan.Fairlie.1977, Oh.Teh.1979.1985, Basler.1984, Baseyan.1979, Matinyan.1981}. A similar solution in $\mbox{O}(4)=\mbox{SU}(2)\otimes \mbox{SU}(2)$ theory \cite{Cervero.1977} and elliptic functions methods in scalar theory were studied \cite{Actor.1979}. Besides, another type of the classical solution in $\mbox{SU}(3)$ Yang-Mills theory described by the Weierstrass elliptic function was found \cite{Tsapalis.2016} and the effects of the elliptic solution on the infrared gluon and on the quantum field theory were discussed \cite{Frasca.2008.2009}.

In the context of the relation with the Higgs particle, the stability of the nonlinear wave solution helps to constrain the mass of the Higgs boson before it was discovered \cite{Achilleos.2012}. In addition, the probability of particle creation process from the vacuum to Higgs particles under the influence of the Jacobi's $dn$ or $1/dn$ functions was calculated \cite{Kitadono.Inagaki.2013}. In the high energy physics or cosmology, the particle creation processes have a special interest. For example, so-called the Schwinger effect \cite{Schwinger.effect} representing the particle creation from the vacuum with a simple uniform and constant electric background field is related to the nonperturbative aspects of the theory. The particle creation in cosmology is important in the context of understanding the history of the universe \cite{Ford.2021}. If we consider the nonlinear massive plane wave solution as the external background field, then what kind of the particle creations will be obtained ? Is there any other nontrivial nature in the consequence of this classical solution ?
The aim of the letter is to answer these questions. As the result, by considering the slightly excited classical solution which will be important when the system is in early universe, we present the nontrivial mass generation and the production probabilities for $W$, $Z$, Higgs and $t$-quark from the vacuum with the nonlinear massive wave solution in the lowest order of perturbation theory.

In this work, we assume that the classical system is in an intermediate state of the phase transition, like  before/after the electroweak phase transition, and it is described by the nonlinear oscillating solution. Moreover, we ignore the effect of expanding universe with a cosmological time scale. If the universe expands, it gives the suppression effect on the oscillation, because the Hubble parameter makes the oscillation energy density diminished (we can find such results for the scalar potential $V(\phi)\propto \phi^n$ with $n=2$ and $n=4$ cases \cite{Turner.1983,Cembranos.2016}). Because the current universe is already expanded enough compared to the early universe, the effect of the nonlinear oscillation in the current universe vanishes effectively. Hence there is no effect on the observed Higgs data at the Large Hadron Collider (LHC).

We briefly review the basic nature of the nonlinear massive solution in the Higgs potential in Sec.~\ref{sec2} for convenience of readers. The particle creation processes for the weak vector bosons, top quark, and the Higgs boson are discussed. In particular, the analytic formulae of the probabilities for these processes per unit spacetime volume are presented in Sec.~\ref{sec3}. Some numerical plots including the probability densities and the velocities of the produced particles are shown in Sec.~\ref{sec4}. Finally, Sec.~\ref{sec5} is devoted to the conclusion of the article.

\section{Nonlinear massive solution in Higgs potential and mass generation \label{sec2}}
To compare the mass generation via the nonlinear massive elliptic solution to the conventional Higgs mechanism, we briefly review how the Higgs mechanism works. The mass generation in the SM is described by the Higgs sector: 
\begin{eqnarray}
	\mathcal{L} 
	= |D_{\mu}\Phi|^2 - V(\Phi) 
	- \lambda_d\bar{Q}_{L} \Phi d_R 
	- \lambda_u\bar{Q}_{L} i\sigma^{2} \Phi^{*} u_R 
	+ \mbox{c.c.},
\end{eqnarray}
where $D_{\mu}$ is the covariant derivative containing $\mbox{SU}_{\rm L}(2)$ and $\mbox{U}_{\rm Y}(1)$ gauge fields, $V(\Phi)=-\mu^2|\Phi|^2 + \lambda|\Phi|^4$ with the negative mass parameter $-\mu^2$ and the self coupling $\lambda$ is the Higgs potential, $\Phi$ is the complex $\mbox{SU}_{\rm L}(2)$ doublet, $\lambda_{u(d)}$ is the Yukawa coupling between the Higgs doublet and $\mbox{SU}_{\rm L}(2)$ left-handed-up(down)-type-field and $\mbox{SU}_{\rm L}(2)$ right-handed field, $\sigma^{2}$ is the Pauli's matrix and $\mbox{c.c}$ stands for the complex conjugate. 

In the unitarity gauge, the $\mbox{SU}_{\rm L}(2)$ complex Higgs doublet field reduces to the form in which only the physical electric-neutral Higgs field remains, i.e., $\Phi(x)=(0,\phi(x)/\sqrt{2})^{t}$, then the Higgs sector which is relevant to the mass generation is given in:
\begin{eqnarray}
\mathcal{L}_{\rm Higgs} 
 &=&    \frac{1}{2}(\partial_{\mu}\phi)(\partial^{\mu}\phi) 
      - V(\phi)
       + \frac{g^2}{4}W^{-\mu}W^{+}_{\mu}\phi^2 
 + \frac{g^2+g^{\prime 2}}{8}Z^{\mu}Z_{\mu}\phi^2 
	  - y_f \bar{\psi}\psi\phi,
	   \label{eq.Higgs.sector}
\end{eqnarray}
where $V(\phi) = -\mu^2\phi^2/2 + \lambda\phi^4/4$, $W^{\pm}_{\mu}$ is the charged $W$-boson and $Z_{\mu}$ is the neutral $Z$-boson, $g$ is the $\mbox{SU}_{\rm L}(2)$ gauge coupling, $g^{\prime}$ is the $\mbox{U}_{\rm Y}(1)$ gauge coupling, and
 $y_f=\lambda_f/\sqrt{2}$ is the scaled Yukawa coupling between a fermion field $\psi$ and the Higgs field $\phi$.  In the conventional Higgs mechanism \cite{Englert.Brout.1964,Higgs.1964,Guralnik.Hagen.Kibble.1964}, the Higgs field $\phi$ possesses the classical vacuum expectation value (VEV) at $\phi=\sqrt{2\mu^2/\lambda}\equiv v$ and the masses for each particle are generated by re-expanding $\phi$ around $v$, namely, $\phi=v+h$ with the excitation of the Higgs field $h$ around the classical vacuum. 
 
The EOM of the self-interacting scalar field in the SM expressed as the following equation 
\begin{eqnarray}
	(\partial_{\mu} \phi(x))(\partial^{\mu}\phi(x)) - \mu^2\phi(x) + \lambda \phi(x)^3 = 0,
\end{eqnarray}
gives the non-linear massive solution described by the Jacobi's elliptic solutions \cite{Actor.1979}. As we discussed in Ref.~\cite{Kitadono.Inagaki.2013}, the Higgs sector in the SM has the nonlinear massive wave solution described by Jacobi's elliptic solution when the classical solutions for all gauge fields vanish. Even in taking into account the Yukawa coupling with a heavy fermion, the SM Higgs sector has the same EOM and the nonlinear-massive-wave solution when all classical-fields vanish except for the Higgs field; namely, $W^{\pm~\mu}_{\rm cl}=0, Z^{\mu}_{\rm cl}=0$ and $\psi_{\rm cl}=0$. The above EOM can be solved by the assuming $\phi_{\rm cl}(x)=\phi_0 dn(p\cdot x, k)$ with the Jacobi's $dn$-type elliptic function \cite{Gradshteyn,Abramowitz}, then we obtain \cite{Kitadono.Inagaki.2013}:
\begin{eqnarray}
	k^2 = 2\left(1-\frac{v^2}{\phi^2_0}\right), 
	\hspace{1cm} p^2 = \frac{\lambda}{2}\phi^2_0,
\end{eqnarray}
where $\phi_0$ is a free parameter which has the mass dimension one. 

\begin{figure}[htb]
	\begin{tabular}{ccc}
		\begin{minipage}[h]{0.35\linewidth}
			\centering
			\includegraphics[scale=0.45]{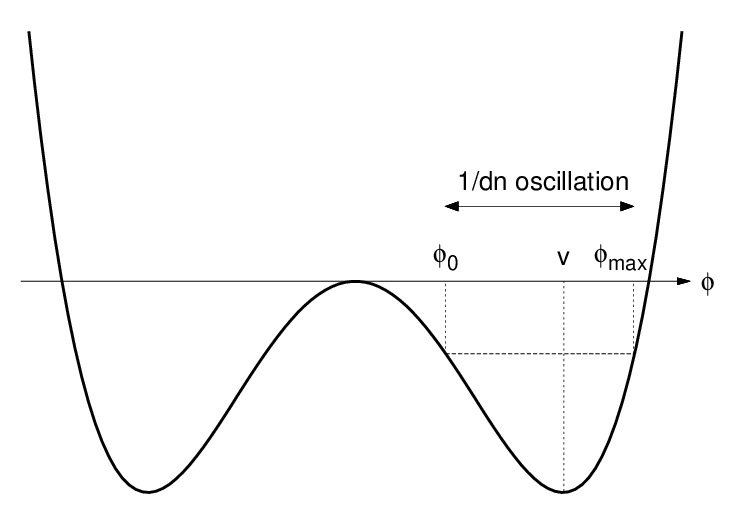}\\
			\hspace{-1.0cm}
			(a) $1/dn$-type
		\end{minipage} 
		&
		\begin{minipage}[h]{0.35\linewidth}
			\centering
	\includegraphics[scale=0.45]{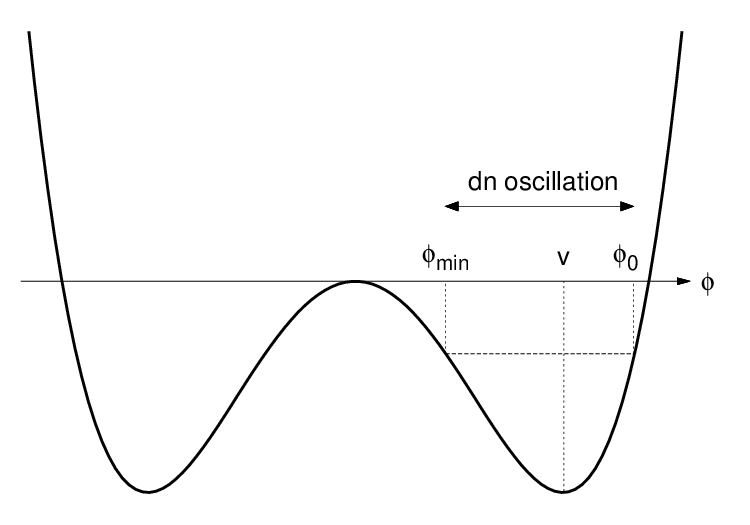}\\
			\hspace{-1.3cm}
			(b) $dn$-type
		\end{minipage}
		\begin{minipage}[h]{0.35\linewidth}
			\centering
	\includegraphics[scale=0.43]{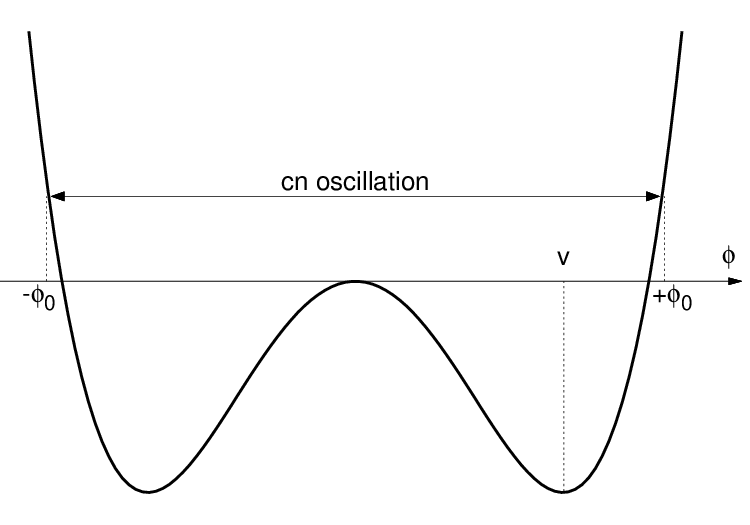}\\
			\hspace{-1.3cm}
			(c) $cn$-type 
		\end{minipage}
	\end{tabular}
	\caption{The three oscillation modes of the classical solution. The maximum (minimum) of the classical field is described by $\phi_{\rm \min(\max)}=v\sqrt{2-\tilde{\phi}^2_0}$ for $1/dn$ ($dn$) oscillation respectively.} 
	\label{fig.potential.three.modes}
\end{figure}

However, it is convenient to rearrange the above solution to avoid imaginary modulus and the expression with $1<k$. Using the mathematical formulae, $dn(u,ik)=1/dn(\sqrt{1+k^2}u,k/\sqrt{1+k^2})$ for $k^2<0$, and $dn(u,k)=cn(ku,1/k)$ for $1<k$ \cite{Gradshteyn, Abramowitz}, after some calculation, the nonlinear massive wave solution can be rewritten in the following form:
\begin{eqnarray}
\phi_{\rm cl}(x)
 = \left\{
 \begin{array}{ll}
 	\frac{\phi_0}{dn(p_1 \cdot x, k_1)} &\mbox{for}~(0 < \tilde{\phi}_0 < 1), \\
 	\phi_0 dn(p_2 \cdot x, k_2) &\mbox{for}~(1 < \tilde{\phi}_0 < \sqrt{2}), \\
 	\phi_0 cn(p_3 \cdot x, k_3) &\mbox{for}~(\sqrt{2} < \tilde{\phi}_0), 
 \end{array}
	\right.
\end{eqnarray}
where $cn$ is the Jacobi's $cn$-type elliptic function, $p_{i}$ and $k_i$ for $i=1,2$ and $3$ are the four-momentum and elliptic modulus for each oscillation mode; (1) $1/dn$-type oscillation for $0< \tilde{\phi}_0 < 1$, (2) $dn$-type oscillation for $1 < \tilde{\phi}_0 < \sqrt{2}$, and (3) $cn$-type oscillation for $\sqrt{2}<\tilde{\phi}_0$ respectively, with the dimensionless initial field value parameter $\tilde{\phi}_0=\phi_0/v$. The invariant mass $\sqrt{p^2_{i}}$ and the elliptic modulus $k_{i}$ for each region ($i=1,2$ and $3$), are explicitly given in
\begin{eqnarray}
\begin{array}{ccc}
\sqrt{p^2_1} &= \frac{M_h}{2} \sqrt{ 2-\tilde{\phi}^2_0 } & \mbox{for}~(0 < \tilde{\phi}_0 < 1), \\
\sqrt{p^2_2} &= \frac{M_h}{2}|\tilde{\phi}_0| &\mbox{for}~(1 < \tilde{\phi}_0 < \sqrt{2}), \\
\sqrt{p^2_3} &= M_h\sqrt{ \frac{\tilde{\phi}^2_0 - 1}{2} }& \mbox{for}~(\sqrt{2} < \tilde{\phi}_0), \label{eq.inv.mass.pcl}
\end{array}
\end{eqnarray}
with $M_H=\sqrt{2\lambda v^2}$ and 
\begin{eqnarray}
\begin{array}{ccc}
	k_1 & = \sqrt{\frac{1-\tilde{\phi}^2_0}{1-\frac{\tilde{\phi}^2_0}{2}}}  
 		&\mbox{for}~(0 < \tilde{\phi}_0 < 1), \\
	k_2 & = \sqrt{\frac{2\tilde{\phi}^2_0-2}{\tilde{\phi}^2_0}}
		&\mbox{for}~(1 < \tilde{\phi}_0 < \sqrt{2}),\\ 
	k_3 & = \sqrt{\frac{\tilde{\phi}^2_0}{2\tilde{\phi}^2_0-2}}
	& \mbox{for}~(\sqrt{2} < \tilde{\phi}_0).
\end{array}
\end{eqnarray}
Note that the above solution satisfies conditions, $\phi_{\rm cl}(x=0)=\phi_0$ with $\partial\phi(x)/\partial x^{0}|_{x_0=0}=0$. The physical interpretation of the three oscillation modes are shown in Fig.~\ref{fig.potential.three.modes}. The oscillation mode expressed by either (a) $1/dn$-type or (b) $dn$-type solution describes the nonlinear oscillation around VEV ($\phi=v$), while the oscillation mode with (c) $cn$-type solution describes the nonlinear oscillation around the origin ($\phi=0$). The nonlinear massive solution describes a dynamically oscillating classical solution which is slightly excited from the globally constant solution ($\phi=v$), depending on the value of $\tilde{\phi}_0$. 

\begin{figure}[htb]
	\centering
	\hspace{2cm}
	\includegraphics[scale=0.85]{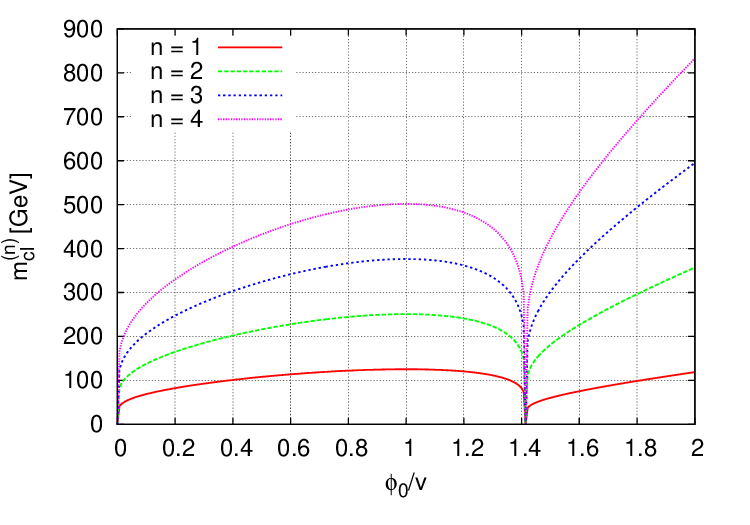}
	\caption{The rest mass of the $n$-th Fourier mode in the classical solution.} 
	\label{fig.dimful.mcln}
\end{figure}

To obtain the probability density of particle creation for weak bosons and fermions, first,  we treat $\phi_{\rm cl}$ as the classical background field and we introduce the quantum fluctuation fields for the massive vector fields $W,Z$ and the massive fermion field $\psi$, respectively. This means that we treat these three fields as the quantized fields. Similarly, we introduce the quantum fluctuation for the scalar field, $h$, through the relation $\phi(x) = \phi_{\rm cl}(x) + h(x)$. Next, we follow the technique of the Fourier series of the elliptic functions used in Ref.~\cite{Kitadono.Inagaki.2013}.
Then, we use the Fourier series expansion for $1/dn$, $dn$, and $cn$ solutions. Using the mathematical formulae \cite{Gradshteyn,Abramowitz}:
\begin{eqnarray}
	\frac{1}{dn(z,k)} = \frac{\pi}{2k^{\prime}K(k)}\left[1+4\displaystyle\sum_{n=1}^{\infty}\frac{(-q)^n}{1+q^{2n}}\cos\left(\frac{n\pi z}{K(k)}\right)\right],\\
	dn(z,k) = \frac{\pi}{2K(k)} + \frac{2\pi}{K(k)}\displaystyle\sum_{n=1}^{\infty}
	\frac{q^n}{1+q^{2n}}\cos\left(\frac{n\pi z}{K(k)}\right),\\
	cn(z,k) =  \frac{2\pi}{kK(k)}\displaystyle\sum_{n=1}^{\infty}
	\frac{q^{n-\frac{1}{2}}}{1+q^{2n-1}}\cos\left(\frac{(2n-1)\pi z}{2K(k)}\right),
\end{eqnarray} 
where the elliptic norm $q$ is given as the function of the elliptic modulus $k$ through the relation,  $q(k)=\exp(-K(k^{\prime})/K(k))$ with $k^{\prime}=\sqrt{1-k^2}$ and $K(k)$ is the complete elliptic integral of the first kind.  
Applying the relation $\cos(z)=(e^{iz}+e^{-iz})/2$ to the above formulae, then we obtain the Fourier series expansion for the classical solution,
\begin{eqnarray}
\phi_{\rm cl}(x) \equiv \phi_{\rm cl,0} + \tilde{\phi}_{\rm cl}(x) 
	 = \phi_{\rm cl,0} + \displaystyle{\sum_{n=-\infty}^{+\infty}} C_n e^{-ip^{(n)}_{\rm cl}\cdot x}, \label{eq.tilde.phicl}
\end{eqnarray}
where the summation over $n$ does not include $n=0$ for $1/dn$ and $dn$ regions, but includes $n=0$ for $cn$ region, and $\phi_{\rm cl,0}$ and $\tilde{\phi}_{\rm cl}(x)$ represent the zero mode and oscillating mode in the classical background field, respectively. The Fourier-zero-mode $\phi_{\rm cl,0}$, the Fourier coefficient $C_n$ and momentum of the $n$-th mode $p^{(n)}_{\rm cl}$ are obtained as:
\begin{eqnarray}
 \phi_{\rm cl,0}
  = \left\{
 \begin{array}{ll}
 	 \frac{\pi \phi_0}{2k^{\prime}_1K(k_1)} &\mbox{for}~(0 < \tilde{\phi}_0 < 1), \\
 	 \frac{\pi\phi_0}{2K(k_2)}              &\mbox{for}~(1 < \tilde{\phi}_0 < \sqrt{2}), \\
 	 0                                      &\mbox{for}~(\sqrt{2} < \tilde{\phi}_0), 
 \end{array}
 \right.
\end{eqnarray}
\begin{eqnarray}
C_n	= \left\{
	\begin{array}{ll}
	  \frac{\pi \phi_0}{k^{\prime}_1K(k_1)} \frac{(-q(k_1))^n}{1+(q(k_1))^{2n}} &\mbox{for}~(0 < \tilde{\phi}_0 < 1), \\
      \frac{\pi \phi_0}{K(k_2)} \frac{(q(k_2))^n}{1+(q(k_2))^{2n}} &\mbox{for}~(1 < \tilde{\phi}_0 < \sqrt{2}), \\
      \frac{\pi \phi_0}{k_3K(k_3)} \frac{(q(k_1))^{n-\frac{1}{2}}}{1+(q(k_3))^{2n-1}}   &\mbox{for}~(\sqrt{2} < \tilde{\phi}_0), 
	\end{array}
	\right.
\end{eqnarray}
and
\begin{eqnarray}
	p^{(n)\mu}_{\rm cl}
	= \left\{
	\begin{array}{ll}
		\frac{n \pi }{K(k_1)}p^{\mu}_{1} &\mbox{for}~(0 < \tilde{\phi}_0 < 1), \\
		\frac{n \pi}{K(k_2)}p^{\mu}_{2}  &\mbox{for}~(1 < \tilde{\phi}_0 < \sqrt{2}), \\
		\frac{(n-\frac{1}{2})\pi}{K(k_3)}p^{\mu}_{3}                                      &\mbox{for}~(\sqrt{2} < \tilde{\phi}_0).
	\end{array}
	\right.
\end{eqnarray}
Furthermore, by squaring the four-momentum $p^{(n)\mu}_{\rm cl}$ and taking its square root, we define the "rest mass" of the $n$-th Fourier mode expressed by $m^{(n)}_{\rm cl}$, as $m^{(n)}_{\rm cl}\equiv \sqrt{(p^{(n)}_{\rm cl})^2}$, then we obtain: 
\begin{eqnarray}
	m^{(n)}_{\rm cl}
=	 \left\{
	\begin{array}{ll}
		\frac{|n| \pi }{K(k_1)} \frac{M_h}{2} \sqrt{2-\tilde{\phi}^2_0} &\mbox{for}~(0 < \tilde{\phi}_0 < 1), \\
		\frac{|n| \pi}{K(k_2)} \frac{M_h}{2}|\tilde{\phi}_0|  
		&\mbox{for}~(1 < \tilde{\phi}_0 < \sqrt{2}), \\
		\frac{|n-\frac{1}{2}|\pi}{K(k_3)} M_h\sqrt{ \frac{\tilde{\phi}^2_0 - 1}{2} }
		&\mbox{for}~(\sqrt{2} < \tilde{\phi}_0),
	\end{array}
	\right.
\end{eqnarray}
where this mass depends on the initial field value parameter $\phi_0$. The numerical plot of $m^{(n)}_{\rm cl}$ for lower $n$ values is shown in Fig.~\ref{fig.dimful.mcln}. 
The values of the masses of the $n$-th Fourier mode in the classical solution increase as $n$ increases, except for the two points at $\tilde{\phi}_0 =0$ and $\tilde{\phi}_0=\sqrt{2}$, because $m^{(n)}_{\rm cl}$ vanishes at these points. 

Because the constant VEV in the Higgs field plays the role of mass in the conventional Higgs mechanism, we can regard the zero mode in the classical background as a kind of "VEV" and arrange the masses of vector bosons and fermion in the Lagrangian. From Eq.~(\ref{eq.Higgs.sector}), the mass and the interactions between the classical background with the quantized-vector-fields are given in:
\begin{eqnarray}
\mathcal{L}_{\rm vector} 
&=& M^2_{W}W^{+}_{\mu}W^{-\mu} + W^{+}_{\mu}W^{-\mu}
(g_{3W}\tilde{\phi}_{\rm cl}+g_{4W}\tilde{\phi}^2_{\rm cl}) \nn\\
&{}&	+\frac{M^2_{Z}}{2} Z_{\mu}Z^{\mu} + Z_{\mu}Z^{\mu}
(g_{3Z}\tilde{\phi}_{\rm cl}+g_{4Z}\tilde{\phi}^2_{\rm cl}),
\end{eqnarray}
where we ignored the irrelevant terms including the kinetic terms for vectors and classical background, etc. One can read each parameter in the Lagrangian easily and the mass of the quantized vector fields and the couplings of the vector fields to the classical field are given in:
\begin{eqnarray}
	M_W &=& \frac{g}{2}\phi_{\rm cl,0}, \hspace{1cm} 
	M_Z = \frac{\sqrt{g^2 + g^{\prime 2}}}{2} \phi_{\rm cl,0}, \\
	g_{3W} &=& \frac{g^2}{2}\phi_{\rm cl,0}, \hspace{1cm} 
	g_{4W} = \frac{g^2}{4}, \\
	g_{3Z} &=& \frac{g^2+g^{\prime 2}}{4}\phi_{\rm cl,0}, \hspace{1cm} 
	g_{4Z} = \frac{g^2+g^{\prime 2}}{8}.
\end{eqnarray}
The mass of the quantized fermion field and the interaction between the classical field and  fermion field are given in:
\begin{eqnarray}
	\mathcal{L}_{\rm fermion}
= -M_f\bar{\psi}\psi - y_f \bar{\psi}\psi\tilde{\phi}_{\rm cl},
\end{eqnarray}
where $M_{f}$ is defined by $M_f = y_f \phi_{\rm cl,0}$. 
The mass of the quantum fluctuation of the Higgs field $h$ around the classical background $\phi_{\rm cl}$ and the interaction between the classical field and the quantized-scalar-field are given in:
\begin{eqnarray}
	\mathcal{L}_{\rm scalar}
	= - M^2_H h^2 + h^2(g_{3H}\tilde{\phi}_{\rm cl} + g_{4H}\tilde{\phi}^2_{\rm cl}),
\end{eqnarray}
where the mass and each coupling are defined by:
\begin{eqnarray}
	M_H = \sqrt{3\lambda}\sqrt{\phi^2_{\rm cl,0} - \frac{v^2}{3}}, \hspace{0.2cm}
	g_{3H} = -3\lambda \phi_{\rm cl, 0}, \hspace{0.2cm}
	g_{4H} = -\frac{3}{2}\lambda. \label{eq.Higgs.mass.coupling}
\end{eqnarray}
It should be noted that the perturbation theory based on the quantized-Higgs-field $h$ looses its validity when $|\phi_{\rm cl,0}|<v/\sqrt{3}$, because the $M_H$ becomes imaginary. 

The heaviest fermion in the SM is the top quark and thus the top Yukawa coupling has the largest value. Hence it will be natural to consider the top quark as the heavy fermion in our calculation. To plot the masses of $W,Z$ vector bosons, Higgs boson and top-quark, we use the numerical values of each coupling as input \cite{PDG,vev.Fermi.muon}:
\begin{eqnarray}
	\lambda &=& 0.13, \hspace{0.5cm} v = 246~\mbox{GeV}, \hspace{0.5cm} y_t = 0.70, \nn\\
	      g &=& 0.653, \hspace{0.5cm} g^{\prime} = 0.351,
\end{eqnarray}
where these values yield the masses of $M_{W}=80.3~\mbox{GeV}$ \cite{WMass1, WMass2}, $M_{Z}=91.2~\mbox{GeV}$ \cite{ZMass1, ZMass2}, $M_{H}=125~\mbox{GeV}$ \cite{HMass1, HMass2}, and $M_{t}=172~\mbox{GeV}$ \cite{TMass1, TMass2}.  

\begin{figure}[htb]
	\centering
	\hspace{2cm}
	\includegraphics[scale=0.85]{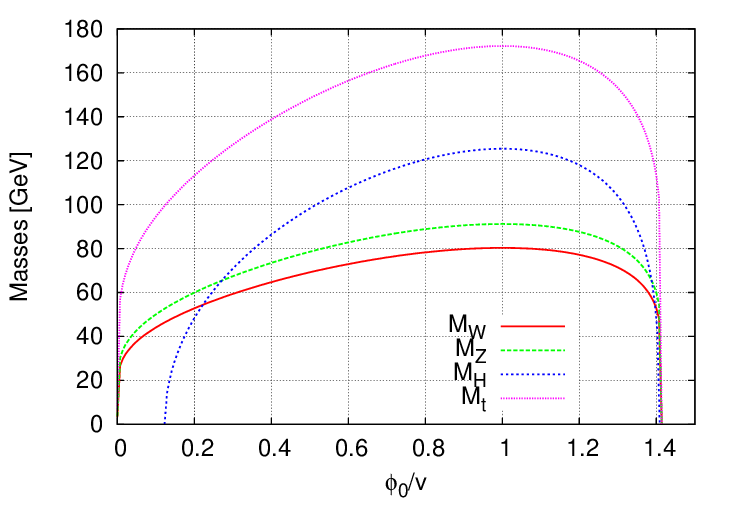}\\
	\caption{The masses for $W,Z$ bosons, Higgs, and $t$-quark in the quantum field theory around the nonlinear massive wave solution.} 
	\label{fig.Meff.all}
\end{figure}

The numerical plots for the masses of $W$-boson, $Z$-boson, Higgs boson and top-quark are shown in Fig.~\ref{fig.Meff.all}.
Basically the masses for all particles are described as the function of $\tilde{\phi}_0$ and all mass reduces to the "standard mass", $M_{W}=80.3~\mbox{GeV}$, $M_{Z}=91.2~\mbox{GeV}$, $M_{H}=125~\mbox{GeV}$, and $M_{t}=172~\mbox{GeV}$, at $\tilde{\phi}_0=1$.  Surprisingly, we can find out a special region that the Higgs mass $M_H$ becomes lightest in the parameter regions, $0.12<\tilde{\phi}_{0}<0.22$ and $1.40<\tilde{\phi}_0<1.41$. This is because that the Higgs mass in Eq.~(\ref{eq.Higgs.mass.coupling}) vanishes at the points where the inside of the square root in the Higgs mass approaches zero, i.e., $\phi_{\rm cl,0} \to v/\sqrt{3}$; while $W,Z$ and $t$-quark masses vanish at the two points where the zero mode approaches zero, i.e., $\phi_{\rm cl,0} \to 0$. Thus the differences of these masses give the interesting mass ordering. 

As is shown in Fig.~\ref{fig.Meff.all}, the conventional Higgs mechanism, "$\mbox{coupling}\times v = \mbox{mass}$", is realized as the special case ($\tilde{\phi}_{0}=1$) of the zero mode in the Fourier series of the dynamical solution. Thus, the masses of particles in the quantum field theory around the nonlinear massive solution is determined by the zero mode $\phi_{\rm cl,0}$ which plays a role of VEV in general case ($\tilde{\phi}_{0}\neq 1$). It is notable that we should not expect these mass spectrum at the current collider physics, for example, at the LHC; because these results only appear during the intermediate era of the phase transition which the system is described by the quantum field theory around the nonlinear massive solution.

\section{Creation of massive particles via nonlinear massive oscillating solution \label{sec3}}
To obtain the probability density for the transition from the vacuum with the classical solution to particles, we use the standard perturbation theory. We take $t\bar{t}$ production process from the vacuum as an example to explain the basic formalism of the computation. 

We consider the transition from the vacuum $|0\rangle$ with the classical background $\tilde{\phi}_{\rm cl}(x)$ to the two body final states of $t\bar{t}$ (top and anti-top) expressed by the state $|\vec{k}_1,\vec{s}_1; \vec{k}_2, \vec{s}_2\rangle$, where $\vec{k}_{1(2)}, \vec{s}_{1(2)}$ are the momentum and spin of the top (anti-top). Then the transition amplitude of the process at the lowest order of the perturbation theory, $A$, is given in
\begin{eqnarray}
	iA = \langle \vec{k}_1,\vec{s}_1; \vec{k}_2, \vec{s}_2 | \left(-iy_t \int d^4x \bar{\psi}_t(x)\psi_t(x)\tilde{\phi}_{\rm cl}(x) \right)|0\rangle,
\end{eqnarray} 
where $\psi_{t}$ and $\bar{\psi}_t$ are the quantized Dirac fields for the top-quark and anti-top-quark, $y_t$ is the top Yukawa coupling, and $\tilde{\phi}_{\rm cl}(x)$ is the classical oscillating solution in Eq.~(\ref{eq.tilde.phicl}). Using the standard quantization procedure with creation/annihilation operators for the Dirac fields \cite{Peskin,Itzykson} and the Fourier series expansion of $\tilde{\phi}_{\rm cl}$ in Eq. (\ref{eq.tilde.phicl}), we obtain
\begin{eqnarray}
	iA = -iy_t \displaystyle\sum_{n}C_n \bar{u}^{s_1}(k_1)v^{s_2}(k_2) (2\pi)^4\delta^4(p^{(n)}_{\rm cl} - k_1 -k_2),
\end{eqnarray}
where $u^{s_1}(k_1) (v^{s_2}(k_2) )$ is $u(v)$-spinor with the momentum $k_1(k_2)$ and spin $s_1(s_2)$. Taking the square of absolute value of the transition amplitude, summing over spins for spinors, dividing the result by the space-time volume $VT=(2\pi)^4\delta^4(0)$, and integrating over the two body phase space for the final particles, we obtain the compact expression for the probability density of the transition per unit spacetime volume, $P(\mbox{vac~with~}\tilde{\phi}_{\rm cl}\to t\bar{t})$:
\begin{eqnarray}
P(\mbox{vac~with~}\tilde{\phi}_{\rm cl}\to t\bar{t}) 
&=& \int \mbox{dPS}_2 \frac{1}{VT}\sum_{s_1,s_2}|A|^2
=
\frac{y^2_t}{2\pi}\displaystyle\sum_{n}C^2_n (\beta^{(n)}_{t})^3 (m^{(n)}_{\rm cl})^2, \label{eq.prob.density.tt}
\end{eqnarray}
where $\mbox{dPS}_2$ represents the two body phase space for $t$ and $\bar{t}$, we took the rest frame of $p^{\mu}_{\rm cl}$ in calculating the phase space integrals, and $\beta^{(n)}_t \equiv \sqrt{1-4M^2_t/(m^{(n)}_{\rm cl})^2}$ is the velocity of the top quark produced from the "decay" of the Fourier $n$-th mode. 

By using the same procedure, we can obtain the similar, but more complicated analytic formula for the transition process from the vacuum to a pair of the vector bosons or a pair of the Higgs particles. The probability density of the $ZZ$ production from the vacuum with the classical solution, $P(\mbox{vac~with~}\tilde{\phi}_{\rm cl} \to ZZ)$, is given in:
\begin{eqnarray}
&& P(\mbox{vac~with~}\tilde{\phi}_{\rm cl}\to ZZ)_{1/dn,dn} \nn\\
&=& \frac{1}{4\pi}
 \left[ \hspace{0.3cm} g^2_{3Z}\displaystyle\sum_{n}C^2_n
 \left\{2 +\left(1-\frac{(m^{(n)}_{\rm cl})^2}{2M^2_Z}\right) \right\}
 \beta^{(n)}_Z \right. \nn\\
&&  \hspace{0.7cm} + \left. 2g_{3Z}g_{4Z}\displaystyle\sum_{m,n}C_{m}C_{n}C_{m+n} 
    \left\{2 +\left(1-\frac{(m^{(m)}_{\rm cl}+m^{(n)}_{\rm cl})^2}{2M^2_Z}\right) \right\}\beta^{(m,n)}_Z \right.\nn\\
&& \hspace{0.7cm} + \left. g^2_{4Z}\displaystyle\sum_{\ell,m,n}C_{\ell}C_{m}C_{n}C_{m+n-\ell}
    \left\{2 +\left(1-\frac{(m^{(m)}_{\rm cl}+m^{(n)}_{\rm cl})^2}{2M^2_Z}\right) \right\}\beta^{(m,n)}_Z
 ~\right], 
 \label{eq.prob.density.ZZ.dndninv}
\end{eqnarray}
where the subscript $1/dn,dn$ stands for that this expression is only valid for $/1dn$ and $dn$ region ($0<\tilde{\phi_0}<\sqrt{2}$);  while the expression of the probability density for $cn$ region ($\sqrt{2}<\tilde{\phi}_0$) is given in
\begin{eqnarray}
  P(\mbox{vac~with~}\tilde{\phi_{\rm cl}}\to ZZ)_{cn} 
 = \frac{g^2_{4Z}}{\pi}
 \displaystyle\sum_{\ell,m,n}C_{\ell}C_{m}C_{n}C_{m+n-\ell},
 \label{eq.prob.density.ZZ.cn}
\end{eqnarray} 
where the two velocities, $\beta^{(n)}_Z$ and $\beta^{(m,n)}_Z$, are defined by
\begin{eqnarray}
\beta^{(n)}_{Z} = \sqrt{1-\frac{4M^2_Z}{(m^{(n)}_{\rm cl})^2}},
\hspace{1.cm}
\beta^{(m,n)}_{Z} = \sqrt{1-\frac{4M^2_Z}{(m^{(m)}_{\rm cl}+m^{(n)}_{\rm cl})^2}}.
\end{eqnarray}
The contribution of $O(g^2_{3Z})$ in Eq.~(\ref{eq.prob.density.ZZ.dndninv}) originates from the interaction $g_{3Z}Z^{\mu}Z_{\mu}\tilde{\phi}_{\rm cl}$, the term of $O(g^2_{4Z})$ comes from the interaction $g_{4Z}Z^{\mu}Z_{\mu}\tilde{\phi}^2_{\rm cl}$, and $O(g_{3Z}g_{4Z})$ is the interference term of the two interactions. The term of $O(g^2_{3Z})$ has the kinematics resembling the "two-body decay" of $n$-th Fourier mode to $ZZ$ and thus this term contains $\beta^{(n)}_Z$ with one integer $n$; while the term of $O(g^2_{4Z})$ has the kinematics resembling the "two-body-transition" from the $m$-th and $n$-th Fourier modes to the $ZZ$ pair and thus this term contains $\beta^{(m,n)}_Z$ with two integers $(m,n)$.

It is noteworthy denoting that the coupling $g_{3Z}$ vanishes and only $g_{4Z}$ contributes to the result in $cn$ region ($\sqrt{2}<\tilde{\phi}_0$), because $g_{3Z}$ is proportional to the zero mode and thus $M_{Z} = 0$ and $\beta^{(n)}_{Z}=\beta^{(m,n)}_{Z}=1$. In addition, the polarization sum for massive vector and massless vector are different, and thus the expressions of the probability density for $1/dn (dn)$ region and $cn$ region are not identical.

Similarly, the probability density of the $W^{+}W^{-}$ production from the vacuum with the classical solution is given in:
\begin{eqnarray}
&&P(\mbox{vac~with~}\tilde{\phi_{\rm cl}}\to W^{+}W^{-}))_{1/dn,dn} \nn\\
&=& \frac{1}{8\pi}
\left[ \hspace{0.3cm} g^2_{3W}\displaystyle\sum_{n}C^2_n
\left\{2 +\left(1-\frac{(m^{(n)}_{\rm cl})^2}{2M^2_W}\right) \right\}
\beta^{(n)}_W \right. \nn\\
&& \hspace{0.7cm} + \left. 2g_{3W}g_{4W}\displaystyle\sum_{m,n}C_{m}C_{n}C_{m+n} 
\left\{2 +\left(1-\frac{(m^{(m)}_{\rm cl}+m^{(n)}_{\rm cl})^2}{2M^2_W}\right) \right\}\beta^{(m,n)}_W 
\right.\nn\\
&& \hspace{0.7cm} 
+ \left. g^2_{4W}\displaystyle\sum_{\ell,m,n}C_{\ell}C_{m}C_{n}C_{m+n-\ell}
\left\{2 +\left(1-\frac{(m^{(m)}_{\rm cl}+m^{(n)}_{\rm cl})^2}{2M^2_W}\right) \right\}\beta^{(m,n)}_W
~\right], 
\label{eq.prob.density.WW.dndninv}
\end{eqnarray}
and
\begin{eqnarray}
 P(\mbox{vac~with~}\tilde{\phi_{\rm cl}}\to W^{+}W^{-})_{cn}
= 
\frac{g^2_{4W}}{2\pi}
\displaystyle\sum_{\ell,m,n}C_{\ell}C_{m}C_{n}C_{m+n-\ell}, \label{eq.prob.density.WW.cn}
\end{eqnarray} 
with the velocities
\begin{eqnarray}
\beta^{(n)}_{W} = \sqrt{1-\frac{4M^2_W}{(m^{(n)}_{\rm cl})^2}},
	\hspace{1.5cm}
\beta^{(m,n)}_{Z} = \sqrt{1-\frac{4M^2_W}{(m^{(m)}_{\rm cl}+m^{(n)}_{\rm cl})^2}}.
\end{eqnarray}

Finally, the interesting creation process of the heavy particle via the nonlinear massive oscillating classical background is the Higgs production. The probability density of the $HH$ (pair of the Higgs particle) production from the vacuum with the classical solution is given in:
\begin{eqnarray}
&& P(\mbox{vac~with~}\tilde{\phi_{\rm cl}}\to HH)_{1/dn,dn}
 \nn\\
&=& \frac{1}{4\pi}
	\left[ \hspace{0.3cm} g^2_{3H}\displaystyle\sum_{n}C^2_n 
	     \beta^{(n)}_H 
+ 2g_{3H}g_{4H}\displaystyle\sum_{m,n}C_{m}C_{n}C_{m+n} 
         \beta^{(m,n)}_H 
+ g^2_{4H}\displaystyle\sum_{\ell,m,n}C_{\ell}C_{m}C_{n}C_{m+n-\ell}
\beta^{(m,n)}_H
~\right], \nn\\
\label{eq.prob.density.HH}
\end{eqnarray}
with 
\begin{eqnarray}
	\beta^{(n)}_{H} = \sqrt{1-\frac{4M^2_H}{(m^{(n)}_{\rm cl})^2}},
	\hspace{1.5cm}
	\beta^{(m,n)}_{H} = \sqrt{1-\frac{4M^2_H}{(m^{(m)}_{\rm cl}+m^{(n)}_{\rm cl})^2}}.
\end{eqnarray}
Note that the contributions of $O(g_{3H}g_{4H})$ and $O(g^2_{4H})$ in Eq.(\ref{eq.prob.density.HH}) were not taken into account in the literature~\cite{Kitadono.Inagaki.2013} and only $O(g^2_{3H})$ contribution was discussed. Thus the above result gives more accurate description than the formula in the literature. On the other hand, the analytic expressions in Eqs.(\ref{eq.prob.density.tt}), (\ref{eq.prob.density.ZZ.dndninv}), (\ref{eq.prob.density.ZZ.cn}), (\ref{eq.prob.density.WW.dndninv}) and (\ref{eq.prob.density.WW.cn}) are newly derived results in this article, up to the best of our knowledge. Note that the above result is only valid for the region where $\phi_{\rm cl,0}>v/\sqrt{3}$ (most of $1/dn$ and $dn$ regions) and we do not extrapolate it to $cn$ region.
\section{Results and discussion \label{sec4}}
In previous section, we have derived the master formulae for the probabilities of the production of $W,Z$ vector bosons, Higgs boson and $t$-quark. Although this formula is expressed as the infinite sum of the Fourier modes, not all Fourier modes can contribute to the probability density. The physical condition on the velocities of the produced particle, i.e., $0\le \mbox{velocity} \le 1$,  tells us which modes can contribute to the probability distribution or not. Some lower Fourier modes do not satisfy this physical condition and we should remove these unphysical modes in numerical computation. 

\begin{figure}[htb]
	\begin{tabular}{ccc}
\hspace{1.cm}
		\begin{minipage}[h]{0.47\linewidth}
			\centering
			\includegraphics[scale=0.65]{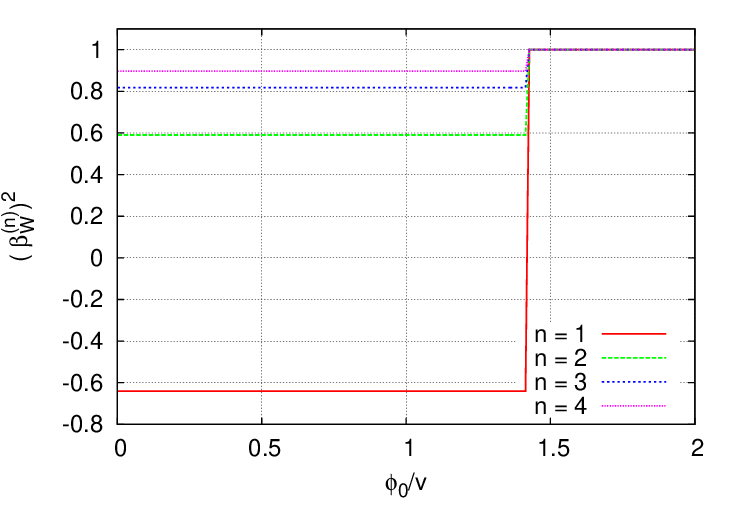}\\
			\vspace{-0.5cm}
			(a) Square of $\beta^{(n)}_W$ \\
			\vspace{0.5cm}
		\end{minipage}
		&
		\begin{minipage}[h]{0.48\linewidth}
			\centering
			\includegraphics[scale=0.65]{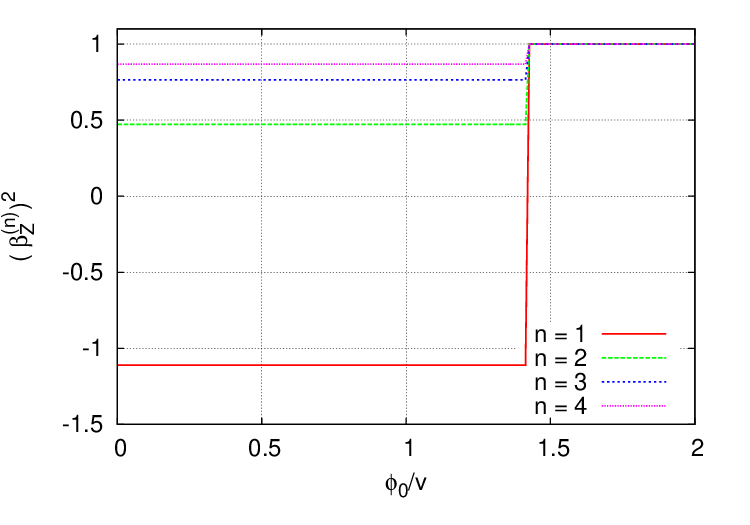}\\
			\vspace{-0.5cm}
			(b) Square of $\beta^{(n)}_Z$ \\
			\vspace{0.5cm}
		\end{minipage}
\\
\vspace{0.5cm}
\hspace{1.cm}
		\begin{minipage}[h]{0.48\linewidth}
		\centering
		\includegraphics[scale=0.65]{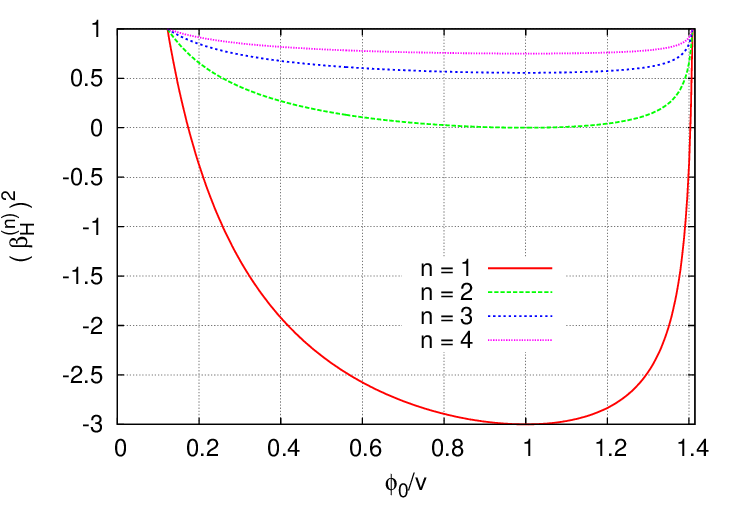}\\
		\vspace{-0.5cm}
		(c) Square of $\beta^{(n)}_H$
	\end{minipage}
	&
	\begin{minipage}[h]{0.48\linewidth}
		\centering
		\includegraphics[scale=0.65]{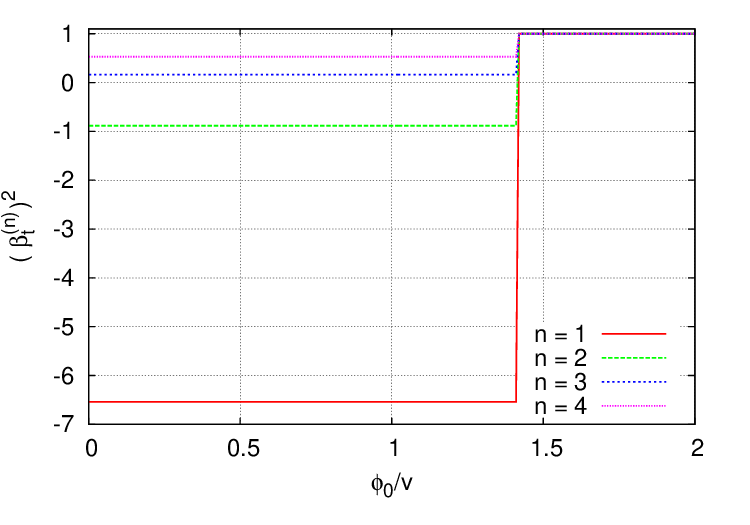}\\
		\vspace{-0.5cm}
		(d) Square of $\beta^{(n)}_{t}$
	\end{minipage}
	\end{tabular}
	\caption{The squared velocities with some lower Fourier modes.} 
	\label{fig.beta.squared.all}
\end{figure}

The squared velocities of $W,Z$ bosons, the Higgs boson, and $t$-quark are shown in Fig.~\ref{fig.beta.squared.all}.
As the plots in Fig.~\ref{fig.beta.squared.all} show the tendencies of the velocities, the velocities of $W,Z$ and $t$-quark indicate that some velocities with lower $n$ do not satisfy the physical condition and thus it does not exist in the "massive region ($1/dn$ and $dn$ regions)", alternatively for $0<\tilde{\phi}_0<\sqrt{2}$, and all velocities reduce to unity in the "massless region ($cn$ region)", alternatively, for $\sqrt{2}<\tilde{\phi}_0$. Obviously, $\beta^{(n \ge 2)}_{W,Z}$ is physical, but $\beta^{(n=1)}_{W,Z}$ is unphysical for $W$ and $Z$ bosons. Similarly, the physical velocity of the top quark is $\beta^{(n\ge 3)}_{t}$, but $\beta^{(n=1,2)}_t$ do not exist. The physical condition for $\beta^{(m,n)}_{W,Z,t}$ is same with $\beta^{(n)}_{W,Z,t}$ and we can use the same discussion for negative $n$. 

The Higgs's velocities are only defined in the restricted massive region due to the uniqueness of the Higgs mass through the self interaction. Interestingly, $\beta^{(n\ge 2)}_H$ is physical for whole massive region and $\beta^{(n=1)}_H$ does not exist for the most of region in $\tilde{\phi}_0$ space, however, $\beta^{(n=1)}_H$ only exists in the limited regions in $\tilde{\phi}_0$ space, namely; 1) $0.123<\tilde{\phi}_0 < 0.174$ and 2) $1.404<\tilde{\phi}_0 < 1.409$.
This $n=1$ contribution characterizes the uniqueness of the probability density for the Higgs production, as we discuss later. 

\begin{figure}[htb]
	\centering
	\hspace{2cm}
\includegraphics[scale=0.85]{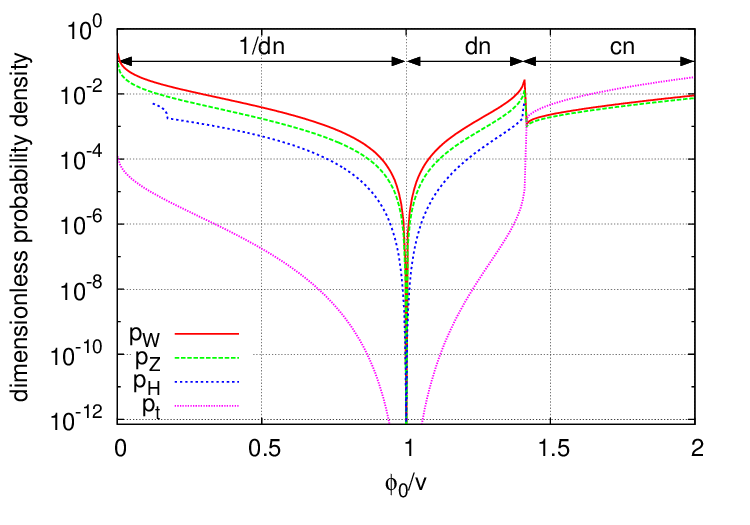}\\
	\caption{The dimensionless probability density for $W^{+}W^{-}$, $ZZ$, $HH$ and $t\bar{t}$.} 
	\label{fig.dimless.prob.density.all}
\end{figure}

To discuss the probability densities for all particles, we define the dimensionless probability density for convenience of numerical plot:
\begin{eqnarray}
	p_{i} \equiv \frac{P(\mbox{vac~with~}\tilde{\phi}_{\rm cl} \to i\bar{i})}{v^4} \hspace{0.5cm} \mbox{for}~(i=W,Z,H,t),
\end{eqnarray}
where $\bar{i}$ indicates the anti-particle of the particle $i$, $p_i$ has no mass dimension and $P$ has the mass dimension four. We plot the dimensionless probability for all particles, $p_{i}$, with $i=W,Z,H$ and $t$ in  Fig.~\ref{fig.dimless.prob.density.all}. 

In $1/dn$ and $dn$ regions ($0<\tilde{\phi}_0<\sqrt{2}$), the probability densities for the production of $W^{+}W^{-}$, $ZZ$, $HH$, and $t\bar{t}$ are described by the solid line (red in color), the dashed line (green in color), the short-dashed line (blue in color), and the dotted line (magenta in color), respectively. Note that all probability densities vanish at $\tilde{\phi}_0=1$, as expected, because the nonlinear massive solution reduces to the constant solution at this point ($\phi_{\rm cl,0}\to v$ and $\tilde{\phi}_{\rm cl}(x) \to 0$), however the probability densities possess non-zero value for $\tilde{\phi}_0 \neq 1$. As we can see, because of the mass ordering, basically $W^{+}W^{-}$ production is the largest, $ZZ$ production probability is the secondly largest, $HH$ production is the thirdly largest, and $t\bar{t}$ production is the lowest one, respectively. 

However, as we can see the "small bump" structure at $\tilde{\phi}_0 \simeq 0.12$ and also at $\tilde{\phi}_0 \simeq 1.40$ (but much narrower than the bump around $\tilde{\phi}_0 \simeq 0.12$), the Higgs production probability is enhanced in these two regions. This bump structure is made of the contribution of the Fourier mode with $n=1$ which only exists in the narrow region in $\tilde{\phi}_0$ space as shown in (c) of Fig.~\ref{fig.beta.squared.all}; while the Fourier modes with $n \ge 2$ exist for whole region in $\tilde{\phi}_0$ space. We point out that this $n=1$ contribution in the Higgs production was missed in the previous literature \cite{Kitadono.Inagaki.2013}. 

On the other hand, in $cn$ region ($\sqrt{2}<\tilde{\phi}_0$), the probability distributions monotonically increase, especially, $t\bar{t}$ production becomes the largest one. Interestingly, all particles $W,Z,t$ behave as the massless particle and the Higgs particle becomes tachyonic and thus we do not plot the probability density for the Higgs in this region. In the massless region, the magnitude of the production probability is governed by the value of the coupling between each particle to the Higgs field and therefore the top production becomes the most important contribution due to the large value of the top Yukawa coupling. 

Besides, the probability densities of the production $W^{+}W^{-}$ and $ZZ$ have the discontinuity at $\tilde{\phi}_0=\sqrt{2}$ due to the decoupling of the zero mode, namely; $g_{3W}$ and $g_{3Z}$ vanish in the $cn$ region, because both $g_{3W}$ and $g_{3Z}$ are proportional to $\phi_{\rm cl,0}$ and the zero mode $\phi_{\rm cl,0}$ vanishes in the $cn$ region. It should be noted again that these transition probability densities in the current universe are negligible and thus it is hard to find any evidence of the probabilities for the processes at the LHC.

Concerning the convergence of the infinite sums of the Fourier modes in the probability densities, we checked numerical convergences in the results. Although the summation over $n$ should be from $-\infty$ to $+\infty$, we terminated the infinite sum at a  large integer $N_{\max}$, namely, we change the sum to $-N_{\max}$ to $+N_{\max}$. According to the numerical check, the first two or three Fourier modes are enough to obtain a good convergence, for example, less than $0.1\%$ for whole regions in $\tilde{\phi}_0$ space except for the two edges $\tilde{\phi}_0=0$ and $\tilde{\phi}_0=\sqrt{2}$. The probability distributions have rapid changes at these two edges. Hence, we chose $N_{\max}$ as a bit higher value, $N_{\max}=7$, in our computation in order to increase the convergence at these two edges. For example, if we change $N_{\max}$ to $8$, the results of $p_t$ at $\tilde{\phi}_0=0.001$ and at $\tilde{\phi}_0=1.414$ change in $3\%$ and $0.3\%$, $p_W$ and $p_Z$ at $\tilde{\phi}_0=0.001$ and at $\tilde{\phi}_0=1.414$ change in $11\%$ and $1\%$, respectively; while $p_H$ at $\tilde{\phi}_0=0.123$ and at $\tilde{\phi}_0=1.408$ change in $10^{-7}\%$. On other points where the results with a quick convergence, for example, at $\tilde{\phi}_0=0.5$, the uncertainty of the convergence is less than $10^{-6}\%$ for all particle. In any case, $N_{\max}=7$ is high enough to give a good convergence in results and the uncertainty is not significant in the logarithmic plot in Fig.~\ref{fig.dimless.prob.density.all}.

\section{Conclusion \label{sec5}}
We revisited the nonlinear massive plan wave solution to discuss the mass generation and particle creations in the quantum field theory around this classical background field.  We studied the probability density distribution of the production of $W$-boson, $Z$-boson, Higgs-boson, and top-quark from the vacuum with the nonlinear massive plane wave background field, as the function of the field value parameter $\phi_0$. Depending on the value of $\phi_{0}$, the Fourier zero-mode in the nonlinear massive solution plays the role of VEV in the Higgs potential and hence the masses of $W$, $Z$, Higgs and top-quark are generated via the analogy of the Higgs mechanism in the region $0<\phi_0 < \sqrt{2}v$; while this zero mode and the masses of each particle vanish in the region $\sqrt{2}v < \phi_0$. 

The oscillating modes in the classical solution triggers the transition of the vacuum to particles which couples to the Higgs field. Hence, we presented the probability densities of the particle production process per unit spacetime for the massive vector bosons, Higgs boson, and top quark. Basically the magnitude of the probability densities is determined by the mass spectrum of the produced particles. However, thanks to the $n=1$ Fourier mode contribution in the Higgs production, the Higgs production has the characteristic behavior in the two regions (small $\phi_0$ region around $\phi_0 \approx 0.12v$ and large $\phi_0$ region around $\phi_0 \approx 1.40v$), which was missed in the previous literature. In addition, the formulae for the production probability densities for $W$-boson, $Z$-boson, and top-quark in $1/dn$ and $dn$ regions and in $cn$ region ($\sqrt{2}v<\phi_0$) are newly presented.

These results will give a new insight into the particle creations in the environment which is slightly excited state from the globally ground state in whole space. If this solution appeared before/after the electroweak phase transition, then the probabilities of the particle production processes for the massive vector bosons, Higgs boson, and heavy fermions, are determined by the formulae presented in the article.

In our computation, we assumed that the physical system is in the slightly-excited state before/after the electroweak phase transition which is described by time-depending nonlinear oscillating solution and we discussed the particle creation processes. Hence the results do not predict any deviation of the observed data at the LHC, because the nonlinear oscillation would be diminished by the effect of the expanding universe. To combine our computations of the generalized mass generation and particle creation probability densities with the cosmological problems, such as the baryon number generation, we must need a treatment based on the non-equilibrium formalism and hence this will be the future direction of the research.

\section*{Acknowledgement}
Y.~K.~was partially supported by the Ministry of Science and Technology of Taiwan (Grant No. MOST 111-2112-M-167-001).

\end{document}